# Towards GaAs Thin-Film Tracking Detectors

**V.-T. Rangel-Kuoppa,**[a,1,*] **S. Ye,**[a,b,*] **Y. J. Noori,**[a,c,*] **W. Holmkvist,**[a] **R. J. Young**[a] **and D. Muenstermann**[a,d,2]

[a] *Department of Physics, Lancaster University, Bailrigg, Lancaster LA1 4YW, UK*
[b] *National Key Laboratory of Science and Technology on Micro/Nano Fabrication, Shanghai Jiaotong University, Shanghai 200240, People's Republic of China*
[c] *School of Electronics and Computer Science, University of Southampton, University Road, Southampton SO17 1BJ, UK*
[d] *Hochschule RheinMain, 65022 Wiesbaden, Germany*

[*] *These authors contributed equally to this work*
  *E-mail:* `tapio.rangel@gmail.com, Daniel.Muenstermann@cern.ch`

ABSTRACT: Silicon-based tracking detectors have been used in several important applications, such as in cancer therapy using particle beams, and for the discovery of new elementary particles at the Large Hadron Collider at CERN. III-V semiconductor materials are an attractive alternative to silicon for this application, as they have some superior physical properties. They could meet the demands for fast timing detectors allowing time-of-flight measurements with ps resolution while being radiation tolerant and cost-efficient. As a material with a larger density, higher atomic number Z and much higher electron mobility than silicon, GaAs exhibits faster signal collection and a larger signal per µm of sensor thickness. In this work, we report on the fabrication of n-in-n GaAs thin-film devices intended to serve next-generation high-energy particle tracking detectors. Molecular beam epitaxy (MBE) was used to grow high-quality GaAs films with doping levels sufficiently low to achieve full depletion for detectors with an active thickness of 10 µm. The signal collection speed of the detector structures was assessed using the transient current technique (TCT). To elucidate the structural properties of the detector, Kelvin probe force microscopy (KPFM) was used, which confirmed the formation of the junction in the detector and revealed residual doping in the intrinsic layer. Our results suggest that GaAs thin films are suitable candidates to achieve thin and radiation-tolerant tracking detectors.

---

[1] Tapio.Rangel@gmail.com
[2] Daniel.Muenstermann@cern.ch

# Contents

## 1. Introduction

Current trends in tracking detectors for high-energy particle physics experiments are moving towards instrumentation of larger and larger areas (e.g. all-semiconductor inner trackers at the experiments at the Large Hadron Collider) and to tracking detectors with timing capability that enable time-of-flight particle identification and/or primary vertex disentanglement in high-pile-up environments.

The use of detectors based on GaAs thin films would be very attractive, as GaAs can provide superb timing properties thanks to its high electron mobility. In addition, the use of binary semiconductors like GaAs and CdTe as solar cells has led to the development of growth techniques with which large areas of thin films can be created at low cost per area, allowing the instrumentation of large areas that would otherwise be financially prohibitive.

GaAs-based detectors would, in addition, have the potential to detect a significant fraction of x-rays and Bremsstrahlung photons: the atomic numbers of Ga and As are $Z^{Ga} = 31$ and $Z^{As} = 33$ respectively, while the atomic number of Si is just $Z^{Si} = 14$. Hence, to detect a 20 keV X-ray photon, the mass attenuation coefficients (μ/ρ) are 42.3 and just 4.4 cm$^2$/g for GaAs and Si, respectively [1]. According to the Beer-Lambert law, the required layer thickness to capture 30% of equivalent ionising particles is 15 μm for GaAs, while more than 300 μm would be required for Si (see figure 1 in [1]).

Using thin films is not only beneficial for minimising the material budget, but it also lowers the distance that the charge carriers have to travel, improving the response time of the detector. Thin-film growth techniques also allow in-situ doping during the growth process, which would save the additional dopant implantation steps in current wafer-based processes. Therefore, epitaxially-grown GaAs film detectors are a good candidate for the next generation of tracking detectors, both in high-energy physics and beyond.

Several studies have investigated GaAs as an alternative material to silicon for radiation detectors, mainly thanks to its properties as room-temperature semiconductor detector. Among the applications that were proposed for these are mammography, where thin-film GaAs detectors could be used to detect the low-energy X-ray photons [2, 3]. GaAs has also attracted interest for tracking detectors in high-energy particle physics experiments [4]. The first studies date back to the 1960s and 1970s reporting alpha-, beta-, gamma- and x-ray detection [5 - 7]. However, at the time of these studies, technological limitations of GaAs growth meant that the performance of the devices produced was greatly limited by impurities and spatial inhomogeneity of the doping levels. In addition, the devices that were proposed in these studies were thick, reducing their performance and cost efficiency compared to the technology available today. In the 1990s and 2000s, studies were published about GaAs thin-film particle detectors [1, 8 - 20], but all these devices used (n+)-type substrates with an intrinsic/low-doped n(-) layer grown on top of it, which was capped with a (p+)-type layer (p(+)-in-n(-)-configuration)[1]. However, this hole-collecting configuration is not compatible with the current pixel read-out chips used in the LHC experiments, e.g. in ATLAS [21, 22]. Its FE-I4 read-out chip requires a detector of either n(+)-in-n(-) or n(+)-in-p(-)-configuration, as the charge collection with electrons in standard silicon pixel detectors is less affected by radiation damage. To the best of our knowledge, no GaAs thin-film particle detector was demonstrated previously with an n-in-n/n-in-p configuration.

The aim of this work was to produce and investigate a GaAs n-in-n/n-in-p thin-film tracking detector as a potential candidate for the ATLAS ITk pixel detector upgrade, which is currently scheduled to be installed by 2026, replacing the commonly used Si tracking detectors. Using a Kelvin probe force microscope (KPFM), a sufficiently low effective doping ($N_{eff}$) level was found for the nominally intrinsic layer, and low leakage currents were observed with current density-voltage (JV) measurements. The performance of these particle detectors was examined using nanosecond laser pulses (Transient Current Technique, TCT) to mimic the incidence of high-energy particles passing through the detector.

This article is divided into the following sections. Following this brief introduction, section 2 gives details of the molecular beam epitaxy (MBE) growth processes used and KPFM measurements performed to analyse the samples' doping levels. Section 3 outlines the processing steps used to create the detector devices using microfabrication techniques. In Section 4, current density-voltage (JV) and capacitance-voltage (CV) measurements are reported and analysed. Section 5 contains the results obtained from performing the TCT measurements on the devices. Finally, our conclusions are reported in Section 6.

## 2. MBE Sample Growth

The n-in-n GaAs detectors for this study were grown on heavily doped (p++)-type GaAs substrates by molecular beam epitaxy (MBE). First, oxygen desorption was performed at 665 °C for 16 min. The substrate temperature was subsequently lowered to $T_{subs}$ = 620 °C, and the following growth procedure was executed: first, a 1 µm p++ GaAs buffer layer (Be-doped) was grown on the (p++) substrate, with a doping level of $10^{18}$ cm$^{-3}$. This was followed by a 10 µm intrinsic (not intentionally doped) layer of GaAs. Finally, a 500 nm n++ GaAs cap layer (Si-doped) was grown on top of the intrinsic layer, with a doping concentration of $10^{18}$ cm$^{-3}$. During the growth, the Ga cell temperature was $T_{Ga}$ = 960 °C while the As cell temperature was $T_{As}$ = 350 °C. The values for the Be (p-type doping) and Si (n-type doping) cells were $T_{Be}$ = 850 °C and $T_{Si}$ = 1200 °C, respectively. All growths were performed at a chamber pressure of ~9×10$^{-9}$ mbar.

It is important to know whether the (not intentionally doped) intrinsic layer has effectively an n- or p-type majority dopant. If the former is the case, the depletion region will be located at the barrier between the buffer layer and the intrinsic layer, while in the latter case, the depletion region will form at the barrier between the intrinsic layer and the cap layer. For operation as a pixel detector, an electrical isolation of the pixels is important to prevent short circuiting between them. The fabrication of a detector with an intrinsic layer of (p-)-type will only require that the pixels are separated by etching through the conductive top (n++) layer. On the other hand, to separate the pixels in case of an effectively (n-) intrinsic layer, it might – depending on the conductivity of the bulk and whether it is possible to obtain "full depletion" (i.e. the depletion zone extending throughout the complete intrinsic layer) – be necessary to etch trenches down to the (p++) buffer layer. The latter case might make the detector fabrication more complicated and would cause the detector to be less efficient, as this reduces the materials that can interact with incident particles. It is therefore preferable to obtain an (n-)-type residual effective doping in the intrinsic layer.

---

[1] "p+" refers to a region p-doped with a "nominal" doping concentration, "p++" to a very highly p-doped region and "p-" to a lowly or non-doped region with residual positive majority charge carriers. The designations apply analogously to n-doped regions.

To identify the majority dopant type of the not-intentionally-doped intrinsic region, the KPFM technique was used. This technique is a variation of the Atomic Force Microscopy (AFM) technique. Briefly, using a conductive tip and a recursive feedback loop, the difference in the work function between the scanned surface and the conductive tip is mapped. A detailed description of this technique can be found in [23]. A piece of the as-grown sample was cut diagonally at approximately 40° from the surface of the epilayer, using the beam exit Ar ion polishing method reported in [24], allowing the KPFM tip to scan through the depth of the sample. A schematic view of the sample after it has been prepared for KPFM measurement is shown in . A KPFM scan is presented in figure 2a showing the KPFM potential vs depth, and in figure 2b, the on-surface electric field obtained using $\vec{E} = -\nabla V$ on the measured data (depicted in figure 2a) is shown.

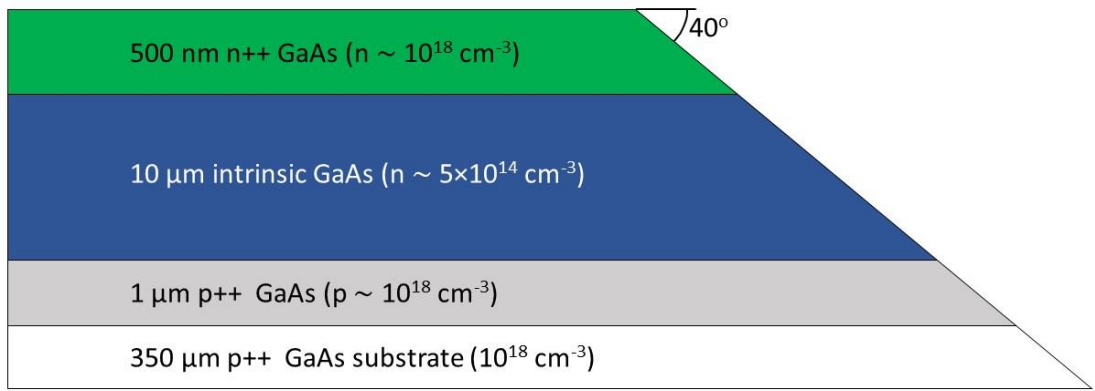

**Figure 1.** Laterally polished sample for Kelvin Probe Force Microscopy (KPFM) measurement. A schematic diagram illustrating a 40° diagonal cut of the sample for KPFM measurement. The GaAs layer doping and thicknesses of each layer are shown.

The data indicates that the change in potential happens between the (p++)-buffer layer and the intrinsic layer, indicating that the intrinsic layer is of (n-)-type doping. This is not ideal, but using the detectors fully depleted (required due to their very shallow active thickness) should still allow the devices to be used with the (n++)-electrodes electrically isolated from each other. Thus, to fabricate pixel detectors, only the top (n++) layer was etched away to isolate the individual pixels. The etch depth was chosen to be 2 μm to form "mesa"-type devices and to be safely inside the intrinsic layer.

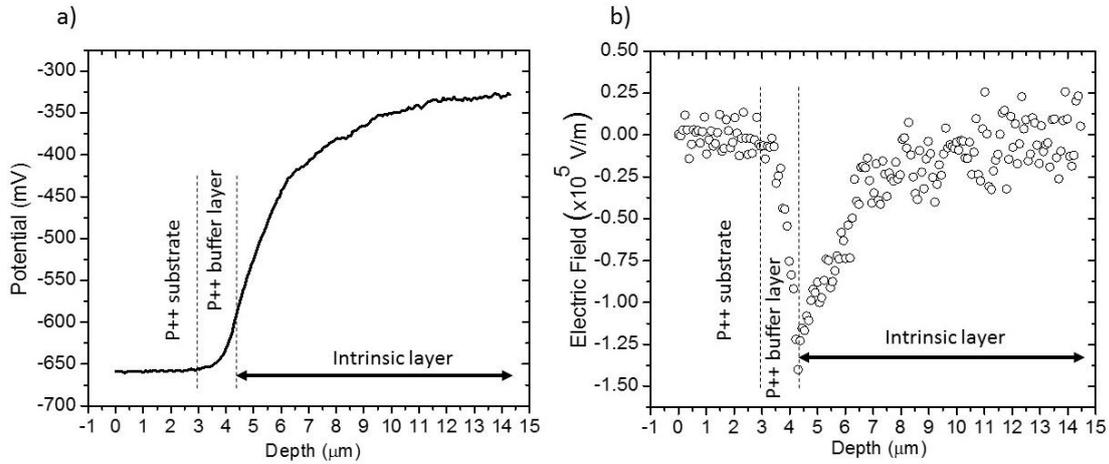

**Figure 2.** Kelvin Probe Force Microscopy (KPFM) measurement and analysis. a) KPFM measurement on the diagonal polished side of the sample shown in figure 1, revealing the Fermi level relative to the bottom of the conduction band as function of depth. b) The on-surface electric field obtained using $\vec{E} = -\nabla V$ on the measured data shown in figure 2a.

## 3. Sample processing

Following MBE growth of the device structure, both pixel detectors and test diodes were fabricated using conventional photolithographic techniques. The processing steps are briefly discussed next and described in figure 3. First, a photoresist was deposited on the sample (see figure 3a), followed by ultraviolet (UV) exposure and developing of the resist (see figure 3b-c) to define the locations of the mesa structures (see figure 3c). The mesa structures were then formed by etching GaAs using a $H_3PO_4:H_2O_2:H_2O$ 1:1:10 solution. The commercialised $H_3PO_4$ has a specified assay of 85% and the commercialised $H_2O_2$ is pre-dissolved in water at 30% (see figure 3d). This solution etches the material at a speed of 500 nm/min [25]. The etching time used was four minutes, to obtain 2 μm high mesas. This mesa height ensures that the mesa is deeper than the 500 nm (n++) cap layer and reaches the intrinsic layer. For the pixel devices, the pixel pitch was chosen to match that of the FE-I4 readout chip, which is 50 μm by 25 μm. For the test diodes, the mesa structures consisted of round circles with diameters of 200, 400 and 800 μm (areas of $A = 31416 \, \mu m^2$, $125664 \, \mu m^2$ and $502655 \, \mu m^2$, respectively). The photoresist was then stripped off (see figure 3e) and an ohmic contact to the back side (i.e. the carrier wafer) was made on the substrate by depositing 10 nm Ni and 100 nm Au in a vacuum chamber (see figure 3f). The sample was then annealed at 410 °C during 5 min in an $N_2$ environment, to achieve the desired Ni-Au alloy as it has been reported in other GaAs-based devices [26, 27]. A second photolithographic process was then performed, by first spin-coating the substrate with a photoresist (see figure 3g), followed by UV-exposure (see figure 3h) and developing steps (see figure 3i), exposing the desired front side to pattern the shape of the top contacts, in "horseshoe-shapes" instead of full-area metal coverage to allow for charge injection by laser pulses. Similarly to the backside contact, Ni and Au layers were then deposited on the front side (see figure 3j). A lift-off process was formed to leave the Ni and Au layers in the desired places. It was then annealed at 410 °C during 5 min in an N$_2$ environment, to make the final device structure (see figure 3k).

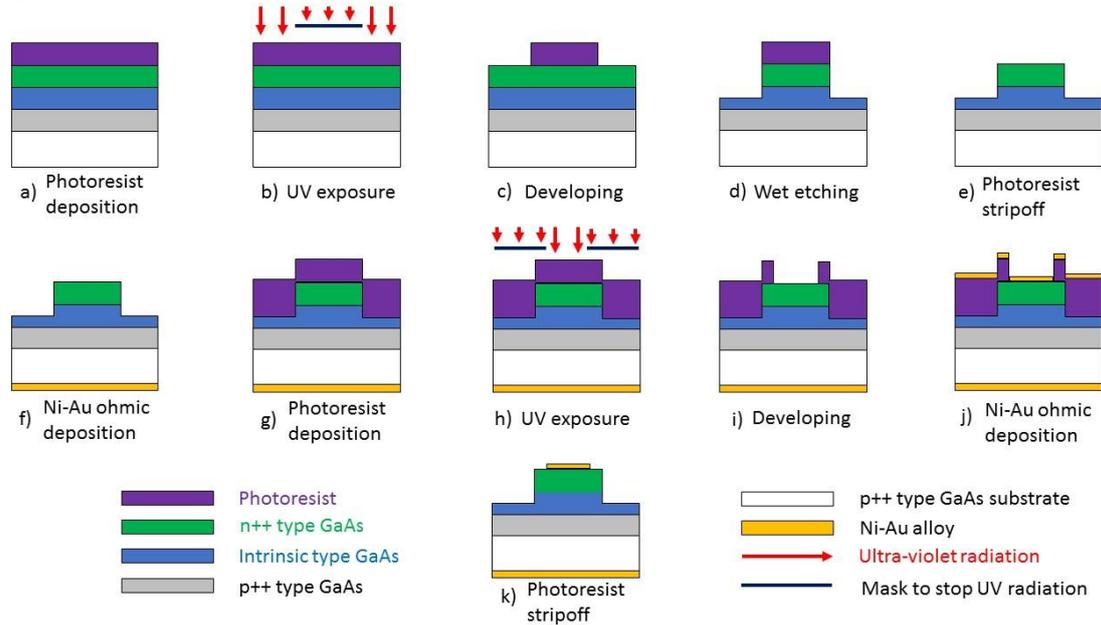

**Figure 3.** Processing steps to achieve the radiation detectors. a) Photoresist deposition using a spin-coater. b) UV exposure on the places where the mesa is desired. c) The photoresist is developed, covering the surface where the mesa will stay. d) Wet-etching is carried out afterwards using a solution of H$_2$O : H$_2$O$_2$ : H$_3$PO$_4$ = 10 : 1 :1 during four minutes, obtaining a 2 µm high mesa. e) Photoresist strip-off. f) The back-side Ni-Au alloy is deposited next and annealed at 410 °C in an N$_2$ environment. g) A second photoresist deposition is done. h) Second UV exposure, this time protecting the photoresist where the front side ohmic contact is desired. i) The photoresist is developed. j) Ni-Au alloy ohmic contact is deposited on the front side and annealed at 410 °C in an N$_2$ environment. k) Finally, the photoresist is stripped off in a lift-off process to obtain the desired sample. The figure is not to scale.

Imaging of the fabricated devices to confirm the dimensions of the final structures was performed using atomic force microscopy (AFM), as shown in figure 4. It is clear from the image that the top metal contact has a thickness of 110 nm and the total mesa depth with a value of 2.1 µm, which falls within good agreement of the desired values. A schematic view of a detector after being processed can be seen in figure 5a. A schematic view of the position of each detector on the sample with the horseshoe-shaped contacts is shown in figure 5b.

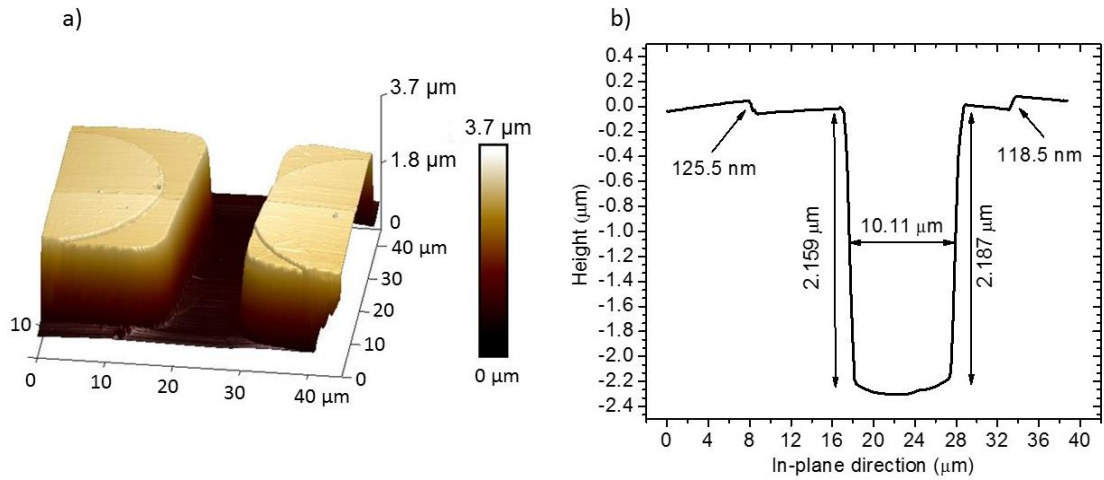

**Figure 4.** Atomic Force Microscopy (AFM) measurement and analysis of a pixel detector. a) Atomic Force Microscopy (AFM) done between two adjacent pixels. b) Lateral height scanning between the two pixels arrays, revealing the desired 2 µm mesa height, together with the 110 nm Ni-Au alloy ohmic

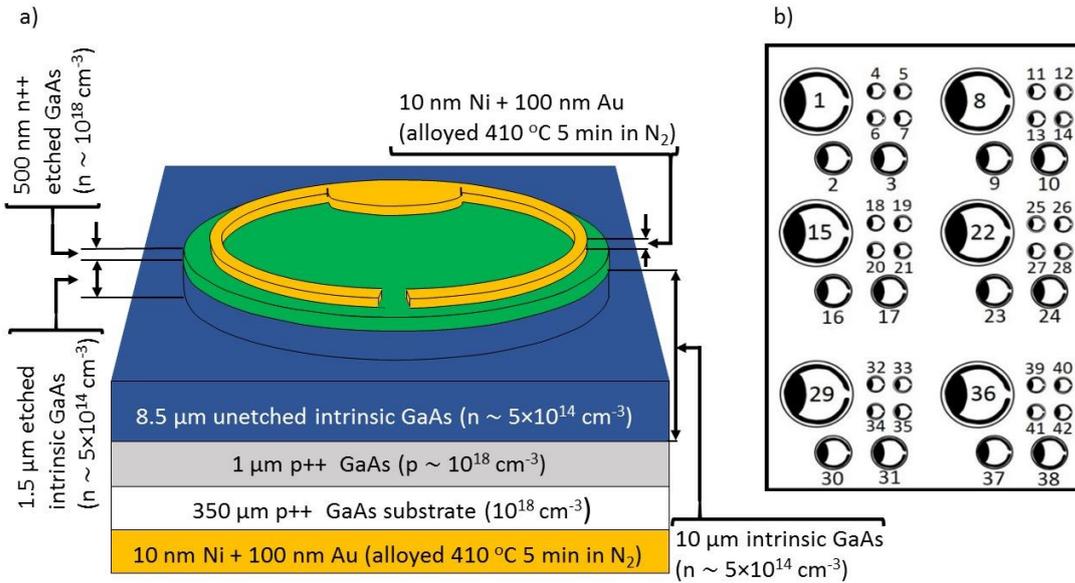

**Figure 5.** Schematic view and arrangement of an array of test diodes. a) Schematic view of a detector after it has been processed, showing the mesa thickness of 2 µm. Also, each layer thickness, doping nature and doping concentration are exposed. b) Schematic view of the position of each detector on the sample, after processing. The outer circle represents the edge of the mesa structure. The numbering of each contact is shown. Contacts 1, 8, 15, 22, 29 and 36 have a mesa diameter of 800 µm, while for contacts 2, 3, 9, 10, 16, 17, 23, 24, 30, 31, 37 and 38 the mesa diameter is 400 µm. The rest of the contacts have a mesa diameter of 200 µm.

## 4. JV and CV measurements

The 42 test diodes shown in figure 5b were characterised using Current Density-Voltage (JV) and Capacitance-Voltage (CV) measurements. A needle probe was placed on each contact while the back contact was connected by placing the sample on an Au-covered plate connected to the other terminal. All measurements were done in darkness and at room temperature.

Contact 1 turned out to be short-circuited and no CV measurement was done to it. All the other contacts showed reasonable and similar JV curves. For clarity purposes, only the most representative absolute JV are shown in figure 6, namely contacts 5, 6, 8, 15, 22, 29 and 36. In the inset the, a linear JV plot is shown for comparison.

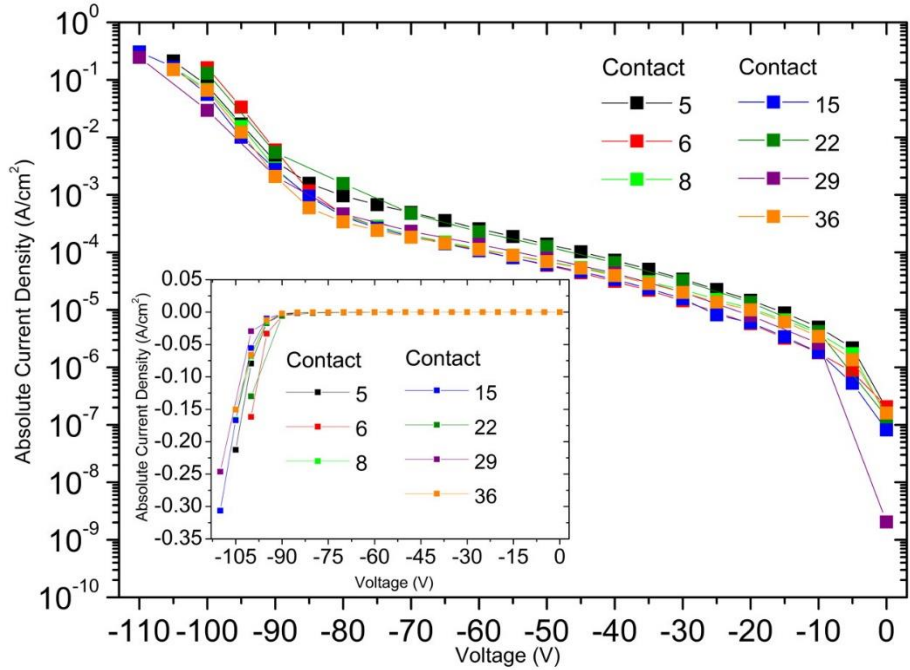

**Figure 6**. Density current versus voltage (JV) measurements. Absolute density current versus voltage (JV) measurements done on darkness and at RT for contacts 5, 6, 8, 15, 22, 29 and 36. In the inset, the linear JV is shown.

As can be seen in figure 6, the break down-voltage (BD) appears to be around -95 V. Discrepancies in the JV could be due to Ga droplets embedded in the samples, formed during Ga "spitting" by the MBE reactor during growth. CV measurements were done at room temperature, in darkness and with a 1 MHz working frequency on contacts 2 to 42: eight contacts reached breakdown voltage before full depletion was achieved. The other 33 thin film detectors (33/42 = 78 %) yield full depletion before reaching BD voltage. For clarity purposes, only the CV measurements on contacts 5, 6, 8, 15, 22, 29 are shown in figure 7, where they are depicted in two separate plots. The $C^{-2}\ vs\ V$ plots are shown in figure 8. It can be observed that full depletion was achieved in contacts 5, 6, 15 and 29 at about -90 V, while at about -80 V, -100 V and -110 V in contacts 36, 8 and 22, respectively.

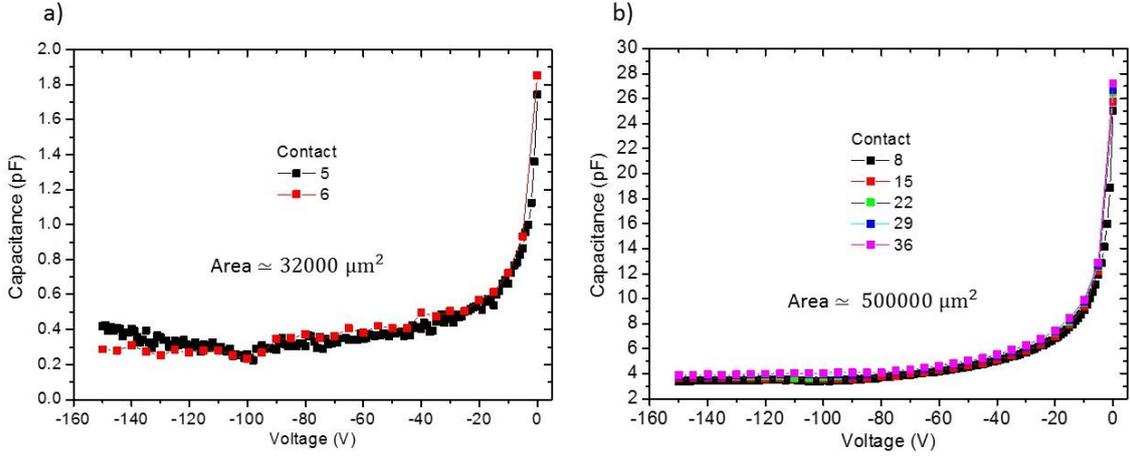

**Figure 7.** Capacitance versus voltage (CV) measurements. a) CV measurements done on darkness and at RT for contacts 5 and 6. b) CV measurements done on darkness and at RT for contacts 8, 15, 22, 29 and 36. The measuring frequency was 1 MHz.

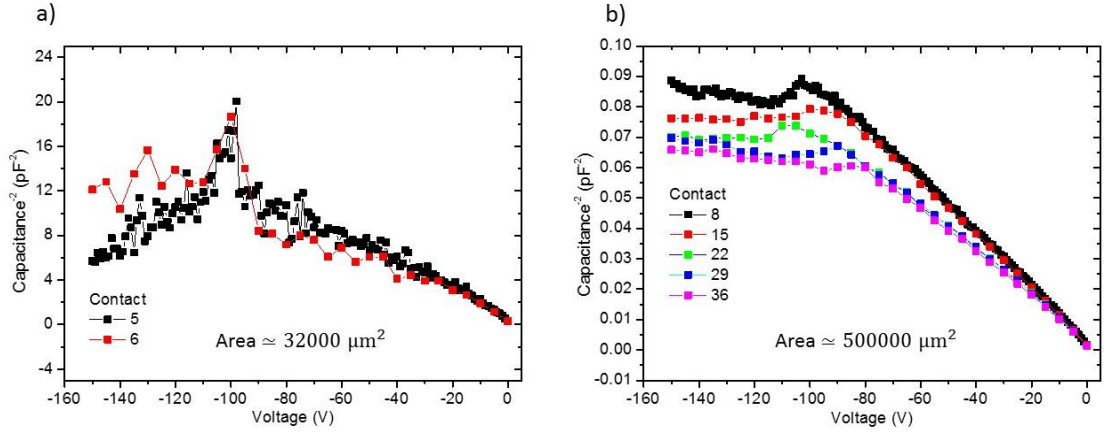

**Figure 8.** Capacitance versus voltage (CV) analysis. a) $C^{-2}$ vs V analysis done on darkness and at RT for contacts 5 and 6. b) $C^{-2}$ vs V analysis done on darkness and at RT for contacts 8, 15, 22, 29 and 36. The measuring frequency was 1 MHz.

The following CV analysis was done to determine the intrinsic (n-type) charge carrier density. The capacitance of a p-n junction is given by [27]

$$C = \frac{\varepsilon A}{d} = A\sqrt{\frac{q\varepsilon}{2(V_{bi}-V)}\frac{N_A N_D}{N_A+N_D}} = A\sqrt{\frac{q\varepsilon}{2(V_{bi}-V)}\left\{\left(\frac{1}{N_A}+\frac{1}{N_D}\right)^{-1}\right\}} \quad (4.1)$$

, where $\varepsilon$ is the high frequency dielectric permittivity of GaAs ($\varepsilon = 96.4\ \frac{pF}{m}$), $d$ the depletion width, $A$ the p-n junction area, $q = 1.6 \times 10^{-19}\ C$ the absolute charge of the electron, $V_{bi}$ the built-in potential, $V$ the applied voltage, $N_A$ the acceptor density and $N_D$ the donor density [27].

In the case when $N_A \gg N_D$ (which is the case in this study) eq. (4.1) is simplified to

$$C = A\sqrt{\frac{q\varepsilon N_D}{2(V_{bi}-V)}} \qquad (4.2)$$

Eq. (4.2) can be rewritten as

$$\frac{1}{C^2} = \left(\frac{2V_{bi}}{N_D q\varepsilon A^2}\right) - \left(\frac{2}{N_D q\varepsilon A^2}\right)V \qquad (4.3)$$

One way to use eq. (4.3) is the following: when linearly fitting eq. (4.3), the slope of the linear fit of $\frac{1}{C^2}$ vs $V$ yields $N_D$. Once $N_D$ has been calculated, then $V_{bi}$ is obtained from the constant of the linear fit. An alternative procedure to use eq. (4.3) is the following: doing the derivative $\frac{\partial}{\partial V}$ in eq. (4.2), and solving for $N_D$, the following expression is obtained

$$N_D = -\left(\frac{2}{q\varepsilon A^2}\right)\left\{\frac{\partial\left(\frac{1}{C^2}\right)}{\partial V}\right\}^{-1} \qquad (4.4)$$

Eq. (4.4), together with (see eq. (4.1))

$$d = \frac{\varepsilon A}{C} \qquad (4.5)$$

allows the charge carrier density to be plotted as a function of the depletion width, i.e., $N_D$ vs $d$. At the same time, it is possible to plot the charge carrier density as a function of the applied voltage $V$, i.e., $N_D$ vs $V$.

The linear fit of eq. (4.3) was done to the CV measurements on contacts 8 and 36 as shown in figure 9, and a charge carrier density of approximately $4 \times 10^{14}\ cm^{-3}$ and $7 \times 10^{14}\ cm^{-3}$, respectively, was obtained in the intrinsic region was obtained. In figure 9b, the donor density vs. depletion width for contacts 8 and 36 is shown, which was obtained using eq. (4.4) and eq. (4.5). It is possible to see the abrupt increase on donor density around 10 µm, confirming full depletion of the intrinsic layer, which is in reasonable agreement with the 10 µm intrinsic layer thickness of the MBE sample.

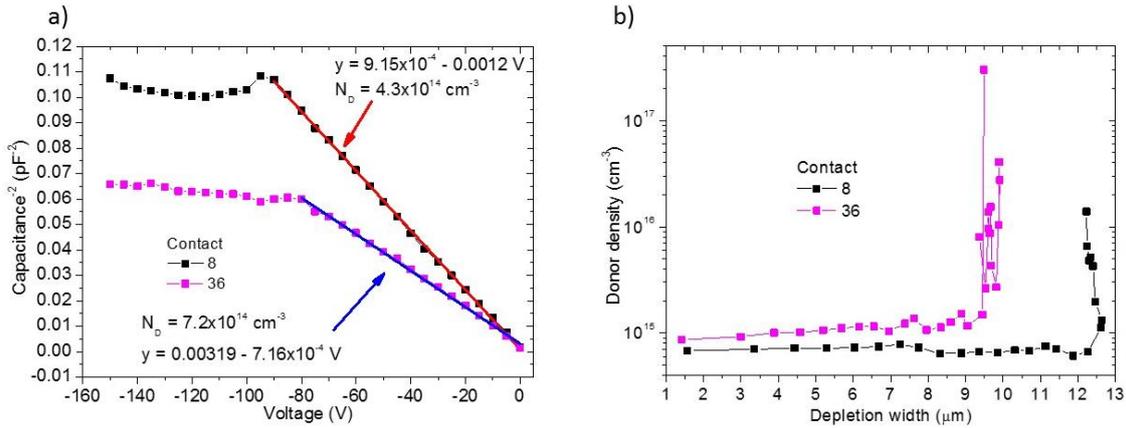

**Figure 9.** Doping concentration calculation and profiling. a) $C^{-2}$ vs V analysis done on CV measurements done on contacts 8 and 36, linearly fitting eq. (4.3). b) Donor density vs depletion width profiling obtained for contacts 8 and 36, using eq. (4.4) and eq. (4.5).

## 5. TCT measurements

The transient current technique (TCT) is a useful technique to obtain information on the electric field and charge carrier behaviour in semiconductor devices. We focused on this technique thanks to its ability to inject consistent amounts of charge carriers, in contrast to ionising radiation, which deposits energy randomly following a Landau distribution. Using TCT, effective trapping times in irradiated / un-irradiated devices can be estimated.

Charge carriers are generated either by laser light or – in exceptional cases – by ionising radiation (usually α-particles), inside a (reverse-)biased detector volume (usually a diode) and are then accelerated towards the respective electrodes by the electric field. The charge movement induces a current in the readout circuit, proportional to the amount of drifting charge and its drift velocity. The current signal shape is also related to the electric field profile that may be affected by lattice defects generated by radiation. Further details can be found in [28].

A schematic view of our TCT setup is depicted in figure 10.

– 10 –

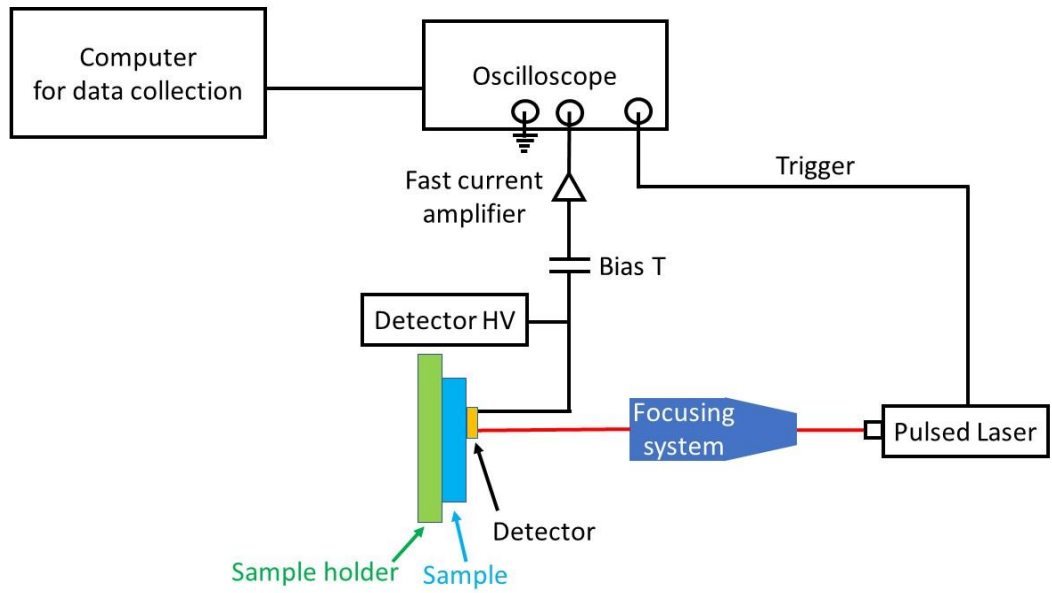

**Figure 10.** Transient current technique (TCT) setup. Schematic view of our experimental TCT setup.

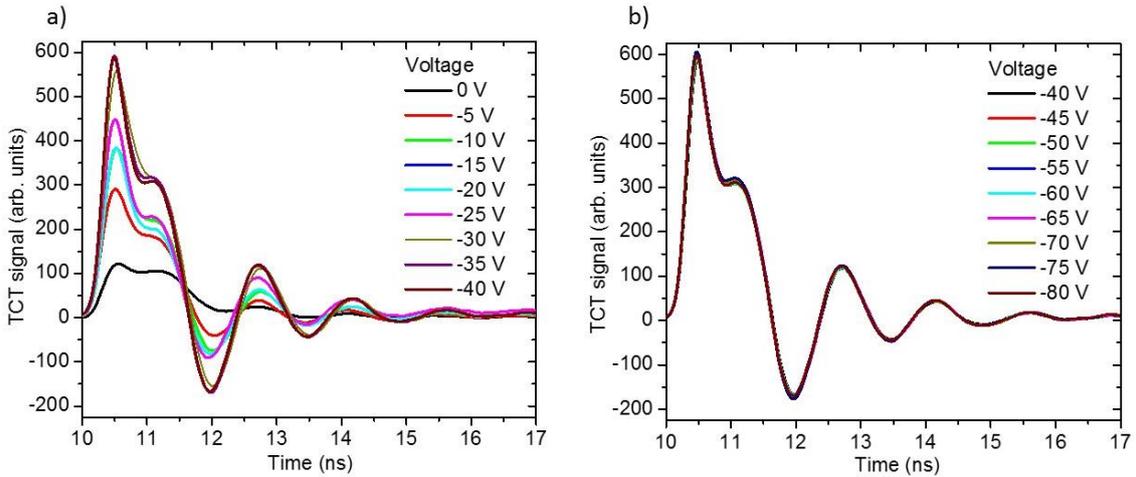

**Figure 11.** Transient Current Technique (TCT) measurements on contact 8. a, TCT measurements done at 0, -5, -10, -15, -20, -25, -30, -35, and -40 V. b, Transient Current Technique (TCT) measurements done at -45, -50, -55, -60, -65, -70, -75 and -80 V.

Contact 8 was investigated via TCT, as it showed the smallest intrinsic charge carrier density. The sample was illuminated using a Fianium SC450 supercontinuum laser, with a wavelength of 680 ± 1 nm, an average power of 1 mW, a repetition rate of 40 MHz and a pulse width of approximately 6 ps. Measurements were done at 0 V, and then at reverse biases of -5 V, -10 V, -15 V, …, -80 V, as shown in figure 11. Clearly, the thin-film detectors react to the laser pulse. The laser pulse ends at the (arbitrary) time of 10 ns, and a clear signal can already be observed at 0V (thanks to diffusion and the built-in voltage), increasing and shortening with increasing reverse bias from 0 V to -40 V. At -40 V, the TCT signal saturates and there is no noticeable change in the TCT signal as the reverse bias increases to -80 V. This suggests that -40 V is already a suitable operation voltage. The signal is extremely short and at the edge of the bandwidth of the current amplifier, but fits the fact that the charge-carrier mobility is very high in GaAs and in addition, the drift length is very short. The large ringing (including the undershoot at 12 ns) are artifacts/reflections due to the high slew rate of the signal.

A two-dimensional scan was done on contact 8, with a reverse bias of -40 V, with the same laser parameters. The sample was scanned in the (x,y)-plane with 10 µm step size, and the result is shown in figure 12. It can be clearly seen that outside the mesa structure, no TCT signal was measured (green colour) confirming the good insulation of the contact via the mesa approach. A very homogeneous signal is obtained within the contact (blue colour), confirming the homogeneous composition of the GaAs sample. This confirms our n-in-n GaAs structure as a promising candidate to achieve thin-film detectors for radiation detection. In summary, the TCT measurement revealed very fast charge collection in GaAs, and homogeneous response to the laser.

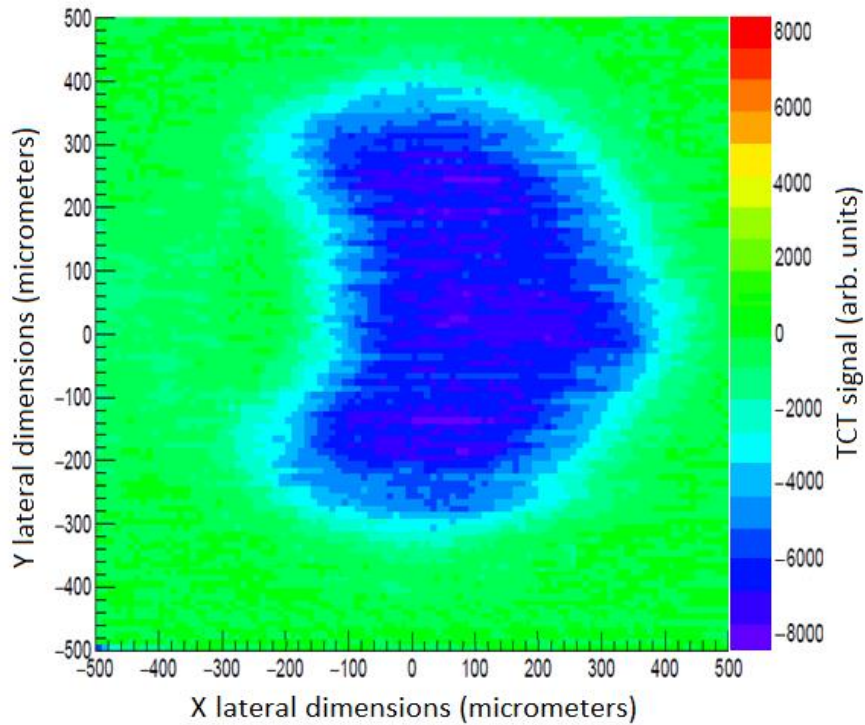

**Figure 12.** Two-dimensional surface scan of transient current technique (TCT) measurements on contact 8. TCT measurements on the sample surface on contact 36. A reverse bias of -40 V was used. Compare to fig. 5 for the expected shape of the electrode area not covered by metal.

## 6. Conclusions

First, n-in-n thin-film detectors were processed from MBE-grown epitaxial GaAs thin films. From an array of 42 test diodes, 33 thin-film detectors have shown full depletion before reaching breakdown voltage, suggesting their suitability for ionising radiation detection, with a respectable 78% (33/42) yield. The JV/CV characterisation was confirmed using the TCT measurement technique. The charge collection time for the thin-film detectors was below 2 ns, close to the bandwidth limitations of the amplifier chain. This superb timing behaviour indicates that GaAs-based thin-film detectors might be excellent candidates for tracking detectors with good timing resolution. In addition, the short charge collection time should provide for excellent radiation tolerance, as the main radiation damage effect, charge trapping, is minimised by fast charge collection. The TCT study revealed a bias voltage of -40 V – below the full depletion voltage – to be a good choice for operational voltage, as larger reverse bias was not found to provide any change in the TCT signal.

With our research, we have shown that it is possible to produce GaAs-based thin-film detectors in an n-in-n configuration, which – to our knowledge – had not been shown. Thanks to intensive research into cost-efficient large substrates for photovoltaic applications, these detectors might be a promising solution to obtain cheaper and thinner radiation detectors, allowing the instrumentation of large areas. We have shown that a 10 µm thick intrinsic GaAs layer, in an n-in-n or n-in-p configuration and yielding about 1100 e-h-pairs for a minimum ionising particle (MIP), might be sufficient for pixel detectors, much less than the usual 150 to 300 µm used in Si-based pixel detectors. Thanks to the fast charge collection time of less than 2ns, such devices can

– 13 –

be expected to be intrinsically radiation-tolerant. Due to the Covid-19 pandemic, it was not possible to obtain measurements with irradiated sensors and using radioactive sources and beam tests, which is left for future research. While MBE-grown epitaxial GaAs thin films can be adapted and are generally of high quality, further research into faster and more cost-efficient processes to grow thin-film detector substrates is required, with photovoltaic applications of thin films having already pioneered this aspect of thin film growth.

## Acknowledgments

Gratitude is expressed to the Newton Fund and Lancaster University. The scientific and technical assistance by Dr Martha San Juan Mucientes and Prof. Oleg Kolosov for AFM and KPFM measurements are thankfully recognized, while the MBE growth was supported by Dr Andrew Marshall, Dr Adam Craig and Dr Laura Hanks, who were invaluable to this research. Finally, for cleanroom support (without which no successful sample processing would have been achieved) we would like to thank Dr Kunal Lulla and Dr Christopher Somerton.